\documentclass{usenix}
\usepackage[noadjust]{cite}
\usepackage{graphicx}
\usepackage{subfigure}
\usepackage{url}

\usepackage[delayed, graphics, dvips, showlabels,sections,textmath,displaymath]{preview}

\newcommand{\beq}{\begin{equation}}
\newcommand{\eeq}{\end{equation}}
\def\bearn{\begin{eqnarray*}}
\def\eearn{\end{eqnarray*}}
\def\barr{\begin{array}}
\def\earr{\end{array}} 

\def\bt{BitTorrent}
\def\p2p{peer-to-peer}
\def\pb{ThePirateBay}

\newcommand{\urlsamefont}[1]
{
\urlstyle{same}\url{#1}
}

\begin{document}
\sloppy

\title{Spying the World from your Laptop\\$-$\\\large{\it{Identifying
     and Profiling Content Providers and Big Downloaders in
     BitTorrent}}}

\author{
\authname{Stevens Le Blond\thanks{This is the author version of the
    paper
published in the Proceedings of the 3rd USENIX Workshop on Large-Scale Exploits and Emergent Threats (LEET'10) in San Jose, CA, on April 27, 2010.}, Arnaud Legout, Fabrice Lefessant, Walid Dabbous, Mohamed Ali Kaafar}
\authaddr{I.N.R.I.A, France}
}

\maketitle 
\sloppy
\begin{abstract}

This paper presents a set of exploits an adversary can use to
continuously spy on most BitTorrent users of the Internet from a
single machine and for a long period of time. Using these exploits for
a period of $103$ days, we collected $148$ million IPs downloading $2$
billion copies of contents. 

We identify the IP address of the content providers for $70\%$ of the
BitTorrent contents we spied on. We show that a few content providers
inject most contents into BitTorrent and that those content providers
are located in foreign data centers. We also show that an adversary
can compromise the privacy of any peer in BitTorrent and identify the
big downloaders that we define as the peers who subscribe to a large
number of contents.  This infringement on users' privacy poses a
significant impediment to the legal adoption of BitTorrent.

\end{abstract}

\section{Introduction}
\label{sec:intro}
\bt{} is one of the most popular \p2p{} (P2P) protocols used today for
content replication. However, to this day, the privacy threats of the
type explored in this paper have been largely
overlooked. Specifically, we show that contrary to common
wisdom \cite{piatek, choffnes, Zhang09tpds_sub}, it is not impractical
to monitor large collections of contents and peers over a continuous
period of time. The ability to do so has obvious implications for the
privacy of BitTorrent users, and so our goal in this work is to raise
awareness of how easy it is to identify not only content provider that
are peers who are the initial source of the content, but also big
downloaders that are peers who subscribe to a large number of
contents.

To provide empirical results that underscore our assertion that one
can routinely collect the IP-to-content mapping on most BitTorrent
users, we report on a study spanning 103 days that was conducted from
a single machine. During the course of this study, we collected 148
million IP addresses downloading $2$
billions copies of contents. We argue that this is a serious privacy
threat for BitTorrent users. Our key contributions are the following.

i) We design an exploit that identify the IP address of the content
  providers for $70\%$ of the new contents injected in BitTorrent.

ii) We profile content providers and show that a few of them inject
  most of the contents in BitTorrent. In particular, the most active
  injects more than 6 new contents every day and are located in
  hosting centers.

iii) We design an exploit to continuously retrieve with time
  the IP-to-content mapping for any peer.

iv) We show that a naive exploitation of the large amount of data
  generated by our exploit would lead to erroneous results. In
  particular, we design a methodology to filter out false positives
  when looking for big downloaders that can be due to NATs, HTTP and
  SOCKS proxies, Tor exit nodes, monitors, and VPNs.

Whereas piracy is the visible part of the lack of privacy in
BitTorrent, privacy issues are not limited to piracy. Indeed,
BitTorrent is provably a very efficient \cite{QIU04_SIGCOMM,
  lego06_IMC} and widely used P2P content replication
protocol. Therefore, it is expected to see an increasing adoption of
BitTorrent for legal use. However, a lack of privacy might be a major
impediment to the legal adoption of BitTorrent. The goal of this paper
is to raise attention on this overlooked issue, and to show how easy
it would be for a knowledgeable adversary to compromise the privacy of
most BitTorrent users of the Internet.

\section{Exploiting the Sources of Public Information}
\label{sec:background}

In this section, we describe the BitTorrent infrastructure and the
sources of public information that we exploit to identify and profile
BitTorrent content providers and the big downloaders.

\subsection{Infrastructure}

At a high level, the BitTorrent infrastructure is composed of three
components: the websites, the trackers, and the peers. The websites
distribute the files containing the meta-data of the contents, i.e.,
.torrent file. The .torrent file contains, for instance, the hostname
of the server, called tracker, that should be contacted to obtain a
subset of the peers downloading that content.

The trackers are servers that maintain the content-to-peers-IP-address mapping
for all the contents they are tracking. Once a peer has downloaded the
.torrent file from a website, it contacts the tracker to subscribe for
that content and the tracker returns a subset of peers that have
previously subscribed for that content. Each peer typically requests
$200$ peers from the tracker every $10$ minutes. Essentially all the
large BitTorrent trackers run the OpenTracker software so designing an
exploit for this software puts the whole BitTorrent community at risk.

Finally, the peers distribute the content, exchange control messages,
and maintain the DHT that is a distributed implementation of the
trackers. 

\subsection{The Content Providers}
\label{sec:info_provider}

BitTorrent content providers are the peers who insert first a content
in BitTorrent. They have a central role because without a content
provide no distribution is possible. We consider that we
  identify a content provider when we retrieve its IP address.  One
  approach for identifying a content provider would be to quickly join
  a newly created torrent and to mark the only one peer with an entire
  copy of the content as the content provider for this
  torrent. However, most BitTorrent clients support the superseeding
  algorithm in which a content provider announces to have only a
  partial copy of the content.  Hence, this naive approach cannot be
  used. In what follows, we show how we exploit two public sources of
  information to aide in identifying the content providers.

\subsubsection{Newly Injected Contents}
\label{sec:info_provider_newly}

The first source of public information that we exploit to identify the
 IP address of the content providers are the websites that list the content that have
just been injected into BitTorrent. Popular websites such as
ThePirateBay and IsoHunt have a webpage dedicated to the newly
injected contents.

A peculiarity of the content provider in a P2P content distribution
network is that he has to be the first one to subscribe to the tracker
in order to distribute a first copy of the content. The webpage of the
newly injected contents may betray that peculiarity because it signals
an adversary that a new content has been injected. \textit{An
  adversary can exploit the newly injected contents to contact the
  tracker at the very beginning of the content distribution and if he
  is alone with a peer, conclude that this peer is the
    content provider.}

To exploit this information, every minute, we download the webpage of
newly injected contents from ThePirateBay website, determine the
contents that have been added since the last minute, contact the
tracker, and monitor the distribution of each content for $24$
hours. If there is a single peer when we join the torrent, we conclude
that this peer is the content provider. We repeated this procedure for
$39,298$ contents for a period of $48$ days from July $8$ to August
$24$, $2009$.

\subsubsection{The Logins}
\label{sec:info_provider_logins}

Sometimes, a content is distributed first among a private community of
users. Therefore, when the content appears in the public community
there will be more than one peer subscribed to the tracker within its
first minute of injection on the website. In that case, exploiting the
newly injected contents is useless and an adversary needs another
source of public information to identify the content provider. The
second source that we exploit are the \textit{logins} of the content providers
on the website. Indeed, content providers need to log
  into web sites using a personal login to announce new
  contents. Those logins are public information.

Moreover, a content provider will often be
the only one peer distributing all the contents uploaded by his
login. The login of a content provider betrays which contents have
been injected by that peer because it is possible to group all the
contents uploaded by the same login on the website. \textit{An
  adversary can exploit the login of a content provider to see whether
  a given IP address is distributing most of the contents injected by
  that login.}

To exploit this information, every minute, we store the login of the
content provider that has uploaded the .torrent file on the webpage of
the newly injected contents. We then group the contents per login and
keep those logins that have uploaded at least $10$ new
contents. Finally, we consider the IP address that is distributing the
largest number of contents uploaded by a given login as the content
provider of those contents. We collected the logins of $6,210$ content
providers who have injected $39,298$ contents for a period of $48$
days from July $8$ to August $24$, $2009$.

We verified that we did not identify the same IP address for many
logins which would indicate that we mistakenly identify an adversary
as content provider. In particular, on $2,206$ such IP addresses, we
identified only $77$ as the content provider for more than $1$ login,
and only $8$ for more than $3$ logins. We performed additional checks
that we extensively describe in Le Blond et al. \cite{angling}.

We validate the accuracy of those two exploits in
Section~\ref{sec:providers_valid} and present their efficiency to
identify the content providers in
Section~\ref{sec:providers_quantify}.

\subsection{The Big Downloaders}
\label{sec:info_big}

For now, we define the big downloaders as the IP addresses that
subscribe to the tracker for the largest number of unique contents. It
is believed to be impractical to identify them because it requires to
spy on a considerable number of BitTorrent users. We now describe the
two sources of public information that we exploit to compromise the
privacy of any peer and to identify the big downloaders.

\subsubsection{Scrape-all: Give Me All the Content Identifiers}

Most trackers support {\it scrape-all} requests for which they return
the identifiers of all the content they track and for each content,
the number of peers that have downloaded a full copy of the content,
the number of peers currently subscribed to the tracker with a full
copy of the content, i.e., seeds, and with a partial copy of the
content, i.e., leechers. A content identifier is a cryptographic hash
derived from .torrent file of a content. Whereas they are not strictly
necessary to the operation of the BitTorrent protocol, scrape-all
requests are used to provide high level statistics on
torrents. \textit{By exploiting the scrape-all requests, an adversary
  can learn the identifiers of all the contents for which he can then
  collect the peers using the announce requests described in
  Section~\ref{sec:info_big_announce}.}

To exploit this information, every $24$ hours, we send a scrape-all
request to all $8$ ThePirateBay trackers and download about $2$
million identifiers, which represents $120$MB of data per tracker. We
then filter out the contents with less than one leecher and one seed
which leaves us with between $500$ and $750$K contents depending on
the day. We repeated this procedure for $103$ days from May $13$ to
August $23$, 2009. ThePirateBay tracker is by far the largest tracker
with an order of magnitude more peers and contents than the second
biggest tracker \cite{Zhang09tpds_sub}, and it runs the OpenTracker
software therefore we limited ourselves to that tracker.

\subsubsection{Announce: Give Me Some IP Addresses}
\label{sec:info_big_announce}

The {\it announce started/stopped} requests are sent when a peer
starts/stops distributing a content. Upon receiving an announce
started request, the tracker records the peer as distributing the
content, returns a subset of peers, and the number of seeds and
leechers distributing that content. When a peer stops distributing a
content, he sends an announce stopped requests and the tracker
decrements a counter telling how many contents that peer is
distributing. We have observed that trackers generally blacklist a
peer when he distributes around $100$ contents. So an adversary should
send an announce stopped request after each announce started requests
not to get blacklisted. \textit{By exploiting announce started/stopped
  requests for all the identifiers he has collected, an adversary can
  spy on a considerable number of users.}

To exploit this information, every $2$ hours, we repeatedly send
announce started and stopped requests for all the contents of
ThePirateBay trackers so that we collect the IP address for at least $90\%$ of the peers
distributing each content. We do this by sending announce started and
stopped requests until we have collected a number of unique
IP addresses
equal to $90\%$ of the number of seeds and leechers returned by the
tracker. This procedure takes around $30$ minutes for between $500$K
and $750$K contents. By repeating this procedure for $103$ days from
May $13$ to August $23$, $2009$, we collected $148$ million IP
addresses downloading $2$ billion copies of contents.

We will see in Section~\ref{sec:down_middle} that once an adversary
has collected the IP-to-content mappings for a considerable number of
BitTorrent users, it is still complex to identify the big downloaders
because it requires to filter out the false positives due to
middleboxes such as NATs, IPv6 gateways, proxies, etc. We will also
discuss how an adversary could possibly reduce the number of false
negatives by identifying the big downloaders with dynamic IP
addresses. Finally, we will see that an adversary can also exploit the
DHT to collect the IP-to-content mappings in
Section~\ref{sec:conclusion}.

\subsection{The Torrent Files}
\label{sec:info_torrent}

Once we have identified the IP address for the content providers and big downloaders, we
use the .torrent files to profile them. A .torrent file contains the
hostname of the tracker, the content name, its size, the hash of the
pieces, etc. Without .torrent file, a content identifier is an opaque
hash therefore, an adversary must collect as many .torrent files as
possible to profile BitTorrent users. For instance, an adversary can
use the .torrent files, to determine if the content is likely to be
copyrighted, the volume of unique contents distributed by a content
provider, or the type of content he is distributing. Clearly, .torrent
files must be public for the peers to distribute contents however, it
is surprisingly easy to collect millions of .torrent files within
hours and from a single machine. \textit{By exploiting the .torrent
  files, an adversary can focus his spying on specific keywords and
  profile BitTorrent users.}

To exploit this information, we collected all the .torrent files
available on Mininova and ThePirateBay websites on May $13$,
$2009$. We discovered $1,411,940$ unique .torrent files on Mininova
and $974,980$ on ThePirateBay. The overlap between both website was
only $227,620$ files. Then, from May $13$, to August $24$, $2009$, we
collected the new .torrent files uploaded on the Mininova,
ThePirateBay, and Isohunt websites. Those three websites are the most
popular and as there is generally a lot of redundancy among the
.torrent files hosted by different websites \cite{Zhang09tpds_sub}, we
limit ourselves to those three.

We will discuss the reasons why our measurement was previously thought
as impractical by the related work in Section~\ref{sec:work}. 

\section{The Content Providers}
\label{sec:providers}

In this section, we run the exploits from
Section~\ref{sec:info_provider} in the wild, quantify the content
providers that we identify, and present the results of their
profiling.

\begin{table}[!t]
  \begin{center}
    \small
    \begin{tabular}{|c|c|c|c|}
      \hline
      $|$Alone$|$ & $|$Login$|$ & $|$Alone $\cap$Login$|$ & Accuracy\\
      \hline
      $21,544$ & $15,308$ & $9,243$ & $99.99\%$\\
      \hline
    \end{tabular}
    \caption{Cross-validation of the two exploits. \textnormal{This
        table shows the accuracy of the two exploits to identify the
        same content provider for the same content. \textit{Alone $\cap$
          Login} is the number of contents for which both sources
        identified a content provider. \textit{Accuracy} is the
        percentage of such contents for which both sources identified
        the same content provider.}}
    \label{tab:infohash-success}
  \end{center}
\end{table} 

\subsection{Identifying the Content Providers}

We start by validating the exploits we use to identify the IP address
of the content providers.

\subsubsection{Validating the Exploits}
\label{sec:providers_valid}

In Section~\ref{sec:info_provider}, we described two exploits to
identify the IP address of a content provider. The first exploit is to connect to the
tracker as soon as a new content gets injected and to check whether we
are alone with the content provider ({\it Alone}). The second exploit
is to find the IP address that has injected the largest number of
contents uploaded by a single login (Login). Whereas it makes sense to
use those exploits to identify content providers, it is necessary to
validate how accurate they are.

We validate the accuracy of these exploits in
Table~\ref{tab:infohash-success}. This table shows that 
  for $9,243$ contents, both exploits identified a content
  provider. Moreover, for $99.99\%$ of those contents both exploits
  identified the same IP address as the content provider. Thus, with
  a high probability the same content providers are identified by two
  independent exploits.

\subsubsection{Quantifying the Identified Content Providers}
\label{sec:providers_quantify}

\begin{figure}[!t]
  \centering
  \includegraphics[width=0.725\columnwidth]{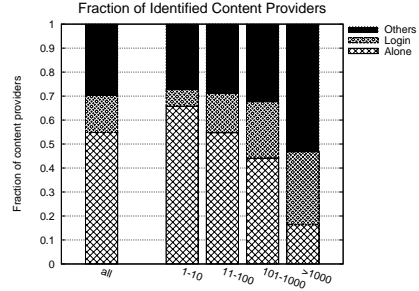}
  \caption{Fraction of content providers that we
    identify. \textnormal{On the x-axis, \textit{all} is for all
      contents, \textit{a-b} is for content with between $a$ and $b$
      peers distributing the content after $24$ hours, and
      \textit{$>1000$} for contents with more than $1,000$ peers
      distributing the content after $24$ hours. \textit{Others} is
      the fraction of content providers that we do not identify.}}
  \label{fig:frac-identified-init-seeds}
\end{figure}

In Fig.~\ref{fig:frac-identified-init-seeds}, we identify
the IP address for $70\%$ of
the content providers injecting $39,298$ new contents over a period of
$48$ days. The fraction of content providers that we identify using
\textit{Alone} only decreases with the number of peers distributing
the content. This is because the more popular the content, the lower
the chances to be alone with the content provider, i.e., from $60\%$
for contents with $10$ peers or less to $17\%$ for contents with more
than $1,000$ peers. However, \textit{Login} compensates for contents
with up to $1,000$ peers. In essence, for contents with more than
$1,000$ peers, we identify close to half of the content providers.

\subsection{Profiling the Content Providers}

\begin{figure}[!t]
  \centering
  \includegraphics[width=0.8\columnwidth,height=0.8\columnwidth]{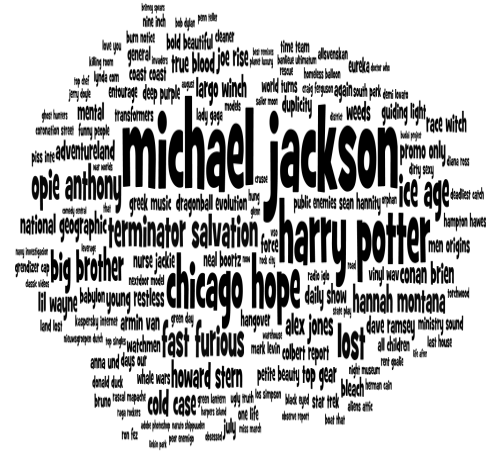}
  \caption{Tag cloud of contents injected by the content providers
    that we have identified. \textnormal{We extract the two most
      significant keywords from each content name contained in the
      .torrent files and vary their police size to reflect the number
      of contents whose name matches those keywords, the largest the
      keywords, the more frequent those keywords appear in the content
      names.}}
  \label{fig:cloudtag-all-init-seed}
\end{figure}

\begin{figure}[!t]
  \centering
  \includegraphics[width=0.725\columnwidth]{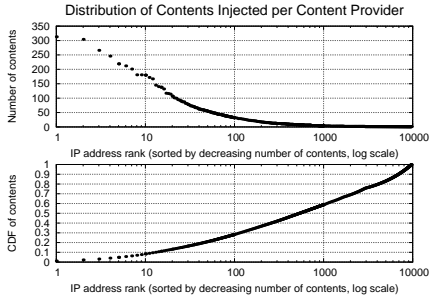}
  \caption{Distribution of the number of contents injected by each
    content provider. \textnormal{The top plot shows the number of
      contents per content provider and the bottom plot shows the CDF
      of contents.}}
  \label{fig:distrib-init-seed}
\end{figure}

We now use the IP address of the content providers that we have
identified for $48$ days to profile their contribution in number of
contents and their location.

\subsubsection{Semantic of the Injected Contents}

Fig.~\ref{fig:cloudtag-all-init-seed} shows a tag cloud of the names
of the contents injected into BitTorrent. This tag cloud suggests that
many contents refer to copyrighted material and that BitTorrent
closely follow events. Indeed, two weeks before we started to identify
the content providers, Michael Jackson died and the latest Happy
Potter movie got released one week after.

\subsubsection{Contribution of the Content Providers}

We see in Fig.~\ref{fig:distrib-init-seed} (top) that some content
providers inject much more contents than others with the most active
injecting more than $300$ contents in $48$ days. The most active
content providers inject more than $6$ contents every day, e.g.,
\textit{eztv} \cite{eztv}, the top content provider, daily injects
$6.5$ TV shows of $430$MB in average. Given the time to capture and
encode a TV show, it suggests that a small community of users injects
contents from the same IP address.

We now look at the contribution of the biggest content providers in
comparison to the total number of injected contents. We see in
Fig.~\ref{fig:distrib-init-seed} (bottom), that the top $100$ content
providers inject $30\%$ of all the contents injected into BitTorrent
and the top $1,000$ content providers inject $60\%$ of all the
contents.

\paragraph{Conclusions}

These results show that few content providers insert most of the
contents. We do not claim that it is easy to stop those content
providers from injecting content into BitTorrent however, it is
striking that such a small number of content providers triggers
billions of downloads. Therefore, it is surprising that the
anti-piracy groups try to stop millions of downloaders instead of a
handful of content providers.

\subsubsection{Location of the Content Providers}

\begin{table}[!t]
  \begin{center}
    \tiny
    \begin{tabular}{|c|c|c|c|c|}
      \hline
      Rank &\# contents & Volume & CC & AS name\\
      \hline
      1 & 313 & 136 & NZ & Vodafone\\
      2 & 304 & 79 & FR & OVH\\
      3 & 266 & 152 & DE & Keyweb\\
      4 & 246 & 34 & FR & OVH\\
      5 & 219 & 186 & FR & OVH\\
      6 & 212 & 247 & DE & Keyweb\\
      7 & 201 & 535 & FR & OVH\\
      8 & 181 & 73 & US & HV\\
      9 & 181 & 17 & CA & Wightman\\
      10 & 180 & 7 & SK & Energotel\\
      11 & 172 & 161 & FR & OVH\\
      12 & 167 & 23 & RU & Corgina\\
      13 & 145 & 197 & DE & Keyweb\\
      14 & 140 & 11 & FR & OVH\\
      15 & 138 & 109 & US & Aaron\\
      16 & 132 & 12 & US & Charter\\
      17 & 117 & 119 & FR & OVH\\
      18 & 116 & 109 & FR & OVH\\
      19 & 114 & 79 & NL & Telfort\\
      20 & 107 & 225 & RU & Matrix\\
      \hline
    \end{tabular}
  \caption{Rank, number of contents, volume of contents (GB), country
    code, and AS name for the top 20 content providers.}
  \label{tab:rank-top20-ips}
  \end{center}
\end{table}

Focusing on the top $20$ content providers in
Table~\ref{tab:rank-top20-ips}, we observe that half of them are
using a machine whose IP address is located in a French and a German hosting center, i.e., OVH and
Keyweb. Those hosting centers provide cheap offers of dedicated
servers with unlimited traffic and a $100$MB/s connection.

However, we observed that the users injecting contents from those
servers are unlikely to be be French or German. Indeed, on $1,515$
contents injected by the content providers from OVH, only $13$
contained the keyword \textit{fr} (French) in their name whereas $552$
contained the keyword \textit{spanish}. Similarly, on $623$ contents
injected from Keyweb, we found $228$ contents with the keyword
\textit{spanish} in their name and none contained the keywords
\textit{fr}, \textit{ge} (German), or \textit{de} (Deutsche). In
conclusion, one cannot easily guess the nationality of a content
provider based on the geolocalization of the IP address
  of the machine he is using to inject contents.

\section{The Big Downloaders}
\label{sec:downloaders}

In this section, we focus on the identification and the profiling of
the big downloaders, i.e., the IP addresses that subscribed in the
largest number of contents. Once we have collected the information
described in Section~\ref{sec:info_big}, it is challenging to identify
and profile the big downloaders because of the volume of
information. Indeed, we collected 148M IP addresses and more than 510M
endpoints (IP:port) during a period of 103 days.

Ordering the IP addresses according to the total number of unique
contents for which they subscribed, we observe a long tail
distribution. In particular, the top $10,000$ IP addresses subscribed
for at least $1,636$ contents and the top $100,000$ IP addresses
subscribed for at least $309$ contents. In the remaining of this
section, we focus on the top $10,000$ IP addresses.

In the following, we show that for many IP addresses, there is a
linear relation between their number of contents and their number of
ports suggesting that those IPs are middleboxes with multiple peers
behind them. However, we will also see that some IP addresses
significantly deviate from this middlebox behavior and we will
identify some of those players with deviant behavior. Finally, we will
profile those players.

\subsection{The Middlebox Behavior}
\label{sec:down_middle}

It is sometimes complex to identify a user based on its IP address or
its endpoint, because the meaning of this information is different
depending on his Internet connectivity. A user can connect through a
large variety of middleboxes such as NATs, IPv6 gateways, proxies,
etc. In all those cases, many users can use the same IP address and
the same user can use a different IP address or endpoints. So an
adversary using the IP addresses or endpoints to identify big
downloaders may erroneously identify a middlebox as a big
downloader. In the following, we aim to filter out those false
positives to identify the big downloaders.

We do not consider false negatives due, for instance, to a big
downloader with a dynamic IP address. It may be possible to identify
big downloaders with a dynamic IP address but it would require a
complex methodology using the port number as the identifier of a user
within an AS; most BitTorrent clients pick a random port number when
they are first executed and then use that port number statically. The
validation of such a methodology is beyond the scope of this paper and
we leave this improvement for future work. However, we will see that
we already find a large variety of big downloaders using public IP
addresses as identifiers.

\begin{figure}[!t]
  \centering
  \includegraphics[width=0.725\columnwidth]{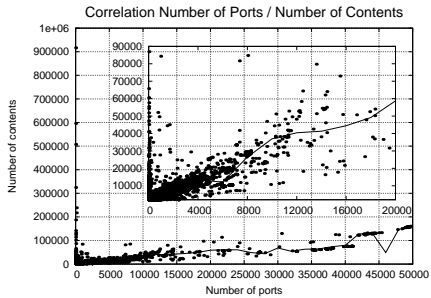}
  \caption{Correlation of the number of ports per IP address and of
    the number of contents for the top $10,000$ IP
    addresses. \textnormal{Each dot represents an IP address. The
      solid line is the average number of contents on the $148$M IP
      addresses computed per interval of $2,000$ ports.}}
  \label{fig:out-corr-port-torrents}
\end{figure}

We confirm the complexity of using an IP address or endpoint to
identify a user in Fig.~\ref{fig:out-corr-port-torrents}. Indeed, we
see that for most of the IP addresses the number of contents increases
linearly with the number of ports. Moreover, the slope of this
increase corresponds to the slope of the average number of contents
per IP over all $148$M IP addresses (solid line). Each new port
corresponds to between $2$ and $3$ additional contents per IP
address. Therefore, it is likely that those IP addresses correspond to
middleboxes with a large number of users behind them. There are also
many IP addresses that significantly deviate from this middlebox
behavior.

\paragraph{Conclusions}
A large number of IP addresses that a naive adversary would classify
as big downloaders actually corresponds to middleboxes such as NATs,
IPv6 gateways, or proxies. However, we also observe many IP addresses
whose behavior significantly deviates from a typical middlebox
behavior.
 
\subsection{Identifying the Big Players}

To understand the role of the IP addresses that deviate from middlebox
behavior, we identify $6$ categories of big players.

\paragraph{HTTP and SOCKS public proxies} 

The two first categories are HTTP and SOCKS public proxies that can be
used by BitTorrent users to hide their IP address from anti-piracy
groups. We retrieved a list of IP addresses of such proxies from the
sites \textit{hidemyass.com} and \textit{proxy.org}. We found $81$
HTTP proxies and $62$ SOCKS proxies within the top $10,000$ IP
addresses.

\paragraph{Tor exit nodes}

The third category is composed of Tor exit nodes that are the outgoing
public interfaces of the Tor anonymity network. To find, the IP
address of the Tor exit nodes, we performed a reverse DNS lookup for
the top $10,000$ IP addresses and extracted all names containing the
\textit{tor} keyword and manually filtered the results to make sure
they are indeed Tor exit nodes. We also retrieved a list of nodes on
the Web site \textit{proxy.org}. We found $174$ Tor exit nodes within
the top $10,000$ IP addresses.

\paragraph{Monitors}
The fourth category is composed of monitors that are peers spying on a
large number of contents without participating in the content
distribution. We identified two ASes, corresponding to hosting centers
located in the US and UK, containing a large number of IP addresses
within the top $10,000$ with the same behavior. Indeed, these IP
addresses always used a single port and we were never able to download
content from them. Therefore, they look like a dedicated monitoring
infrastructure instead of regular peers. We found $1,052$ such IP
addresses within only two ASes in the top $10,000$ IP addresses

\paragraph{VPNs}

The fifth category is composed of VPNs that are SOCKS proxies
requiring authentication and whose communication with BitTorrent users
is encrypted. To find VPNs, we performed a reverse DNS lookup for the
top $10,000$ IP addresses and extracted all names containing the
\textit{itshidden}, \textit{cyberghostvpn}, \textit{peer2me},
\textit{ipredate}, \textit{mullvad}, and \textit{perfect-privacy}
keywords and manually filtered the results to make sure they are
indeed the corresponding VPNs. Those keywords correspond to well-known
VPN services. We found $30$ VPNs within the top $10,000$ IP addresses.

\paragraph{Big downloaders}

The last category is composed of big downloaders that we redefine as
the IP addresses that \textit{distribute} the largest number of
contents and that are used by a few users. We selected the IP
addresses we could download content from and that used fewer than $10$
ports. Hence, those IP addresses cannot be a monitors as we downloaded
content from them and they cannot be large middleboxes due to the
small number of ports. We found $77$ such big downloaders.

\paragraph{Conclusions}

We have identified $6$ categories of big players including the big
downloaders. We do not claim that we have identified all categories of
players nor found all the IP addresses that belong to one of those $6$
categories. Instead, we have identified few IP addresses in each
category within the top $10,000$ peers that we use in the following to
profile the big players.

\begin{figure}[!t]
  \centering
  \includegraphics[width=0.725\columnwidth]{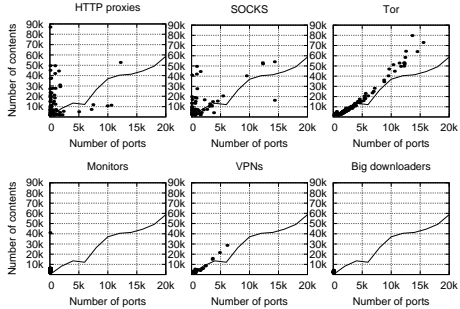}
  \caption{Correlation of the number of ports per IP address and of
    the number of contents of the big players. \textnormal{Each dot
      represents an IP address. The solid line represents the
      middlebox behavior.}}
  \label{fig:out-actors-footprint}
\end{figure}

\subsection{Profiling the Big Players}

\begin{figure}[!t]
  \centering
  \includegraphics[width=0.725\columnwidth]{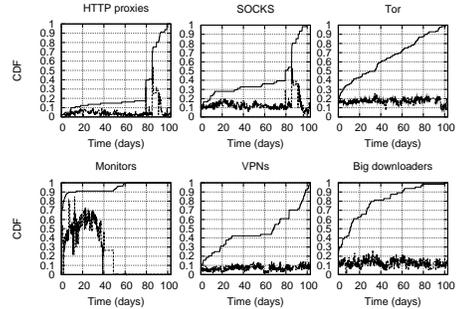}
  \caption{Activity of the big players in time. \textnormal{For each
      category, the dashed line represents the fraction of the top
      $10,000$ IP addresses of a given snapshot that belongs to the
      top $10,000$ IP addresses on all snapshots. The solid line
      represents, for each category, the fraction of the top $10,000$
      IP addresses on all previous snapshots that belongs to the top
      $10,000$ IP addresses on all snapshots.}}
  \label{fig:out-actors-benef-cumul-crawl}
\end{figure}

We see in Fig.~\ref{fig:out-actors-footprint} that for HTTP and SOCKS
proxies the number of contents per IP address is much larger than for
middleboxes (solid line). Considering the huge number of contents
these IP addresses subscribed to, it is likely that the proxies are
used by anti-piracy groups. Indeed, we see in
Fig.~\ref{fig:out-actors-benef-cumul-crawl} that our measurement
system suddenly stops seeing the IP addresses of monitors after day
$50$. In fact, by that date, ThePirateBay tracker changed its
blacklisting strategy to reject IP addresses that are subscribed to a
large number of contents. Whereas it was not a problem for our
measurement system because it uses announce stopped requests as
described in Section~\ref{sec:info_big_announce}, monitors got
blacklisted. However, we observe on day $80$ that the number of HTTP
and SOCKS proxies suddenly increased, probably corresponding to
anti-piracy groups migrating their monitoring infrastructure from
dedicated hosting centers to proxies. Considering, the synchronization
we observe in Fig.~\ref{fig:out-actors-benef-cumul-crawl} in the
activity of the HTTP and SOCKS proxies, it is likely that those
proxies were used in a coordinated effort.

The correlation for monitors and big downloaders in
Fig.~\ref{fig:out-actors-footprint} does not show any striking result,
therefore we do not discuss it further. However, we observe in
Fig.~\ref{fig:out-actors-footprint} that for Tor exit nodes and VPNs
the number of contents per IP address is close to the IP addresses of
the middleboxes (solid line). For large number of ports, Tor exit
nodes deviate from the standard middlebox behavior. In fact, we found
that just a few IP addresses are responsible of this deviation, all
other Tor exit nodes following the trend of the solid line. We believe
that those few IP addresses responsible for the deviation are used by
either big downloaders or anti-piracy groups.

\paragraph{Conclusions}

We have shown that many peers do not correspond to a BitTorrent user
but to monitors or to middleboxes with multiple users behind
them. These peers introduce a lot of noise for an adversary who would
like to spy on BitTorrent users and in particular on the big
downloaders. However, we have shown that it is possible to filter out
that noise to identify the IP address and profile the big downloaders.

\section{Related Work}
\label{sec:work}

As far as we know, no related work has explored the identification of
the content providers in BitTorrent so both the data and the results
concerning these players are entirely new.

Some related work has measured BitTorrent at a moderate scale but none
at a large-enough scale to identify the big downloaders. This is
because most of the measurements inherited two problems from using
existing BitTorrent clients \cite{SiganosPam09, piatek2, piatek}. The
first problem is that existing clients introduce a huge computational
overhead on the measurement. For instance, each announce started
request takes one fork and one exec. Therefore, the measurement is
hard to efficiently parallelize.

The second problem is that regular BitTorrent clients do not exploit
all the public sources of information that we have presented in
Section~\ref{sec:info_big} and \ref{sec:info_torrent}. A content
identifier is essentially the hash of a .torrent file. So not
exploiting scrape-all requests limits the number of spied contents to
the number of .torrent files an adversary has collected. In addition,
clients may not be stopped properly and so not send the announce
stopped request, making the measurement prone to blacklisting.

In the following, we describe how the scale of previous measurements
differs from ours according to the sources of public information that
they exploit.

\subsection{No Exploitation of Scrape-all Requests}

We split the related work not exploiting scrape-all requests into two
families: A first family spying on few contents and a second one using
a large infrastructure to spy on more contents. Siganos et
al. measured the top $600$ contents from The \pb{} \cite{SiganosPam09}
Web site during $45$ days collecting $37$ million IP addresses. Using
only the top $600$ contents does not allow an adversary to identify
the big downloaders. The same remark holds for Choffnes et
al. \cite{choffnes} who monitored $10,000$ peers and did not record
information identifying contents therefore they cannot either identify
the big downloaders.

The second family spied on more contents but using a large
infrastructure. Piatek et al. used a cluster of workstations to
collect $12$ million IP addresses distributing $55,523$ contents in
total \cite{piatek, piatek2}. It is unclear how many simultaneous
contents they spied as they reported being blacklisted when being too
aggressive, suggesting that they did not properly send announce
stopped requests.

Finally, Zhang et al. \cite{Zhang09tpds_sub} is the work that is the
closest to ours in scale however, they used an infrastructure of $35$
machines to collect $5$ million IP addresses within a $12$ hours
window. In comparison, our customized measurement system used $1$
machine to collect around $7$ million IP addresses within the same
time window, making it about $50$ times more efficient. In addition,
that we performed our measurement from a single machine demonstrates
that virtually \textit{anyone} can spy on BitTorrent users, which is a
serious privacy issue.

\subsection{No Exploitation of Announce Requests}

Dan et al. measured $2.4$ million torrents with $37$ million peers,
but used a different terminology \cite{dan}. Indeed, they performed
\textit{only} scrape-all requests so they knew the number of peers per
torrent but not the IP addresses of those peers. This data is much
easier to get and completely different in focus.

\section{Discussion and Conclusions}
\label{sec:conclusion}

We have shown that enough information is available publicly in
BitTorrent for an adversary to spy on most BitTorrent users of the
Internet from a single machine. At any moment in time for $103$ days,
we were spying on the distribution of between $500$ and $750$K
contents. In total, we collected $148$M of IP addresses distributing
$1.2$M contents, which represents $2$ billion copies of content.

Leveraging on this measurement, we were able to identify 
  the IP address of the content
providers for $70\%$ of the new contents injected into BitTorrent and
to profile them. In particular, we have shown that a few content
providers inject most of the contents into BitTorrent making us wonder
why anti-piracy groups targeted random users instead. We also showed
that an adversary can compromise the privacy of any peer in BitTorrent
and identify the IP address of the big downloaders. We have seen that it was complex to
filter out false positives of big downloaders such as monitors and
middleboxes and proposed a methodology to do so.

We argue that this privacy threat is a fundamental problem of open P2P
infrastructures. Even though we did not present it in this paper, we
have also exploited the DHT to collect IP-to-content mappings using a
similar methodology as for the trackers. That we were also able to
collect the IP-to-content mappings on a completely different
infrastructure reinforces our claim that the problem of privacy is
inherent to open P2P infrastructures.

A solution to protect the privacy of BitTorrent users might be to use
proxies or anonymity networks such as Tor, however a recent work shows
that it is even possible to collect the IP-to-content mappings of
BitTorrent users on Tor \cite{btorNsdiPoster}. Therefore, the degree to which it
is possible to protect the IP-to-content mappings of P2P filesharing
users remains an open question.

\paragraph*{Acknowledgments}
We would like to thank Thierry Parmentelat and T. Bar\i{}\c s Metin
for their system support and the anonymous reviewers for their useful
comments.
\begin{small}

\end{small}

\end{document}